# Prediction of Dashed Cafe Wall illusion by the Classical Receptive Field Model

Nasim Nematzadeh
College of Science and Engineering
Flinders University
Adelaide, Australia
nasim.nematzadeh@flinders.edu.au

David M. W. Powers
College of Science and Engineering
Flinders University
Adelaide, Australia
david.powers@flinders.edu.au

*Abstract*— The Café Wall illusion is one of a class of tilt illusions where lines that are parallel appear to be tilted. We demonstrate that a simple Differences of Gaussian model provides an explanatory mechanism for the illusory tilt perceived in a family of Café Wall illusion generalizes to the dashed versions of Café Wall. Our explanation models the visual mechanisms in low-level stages and the lateral inhibition of simple cells that can reveal tilt cues in Geometrical distortion illusions such as Tile illusions particularly Café Wall illusions. For this, we simulate the activations of the retinal/cortical simple cells in responses to these patterns based on a Classical Receptive Field (CRF) model (refered to as Vis-CRF) to explain tilt effects in these illusions. Previously, it was assumed that all these visual experiences of tilt arise from the orientation selectivity properties described for more complex cortical cells. An estimation of an overall tilt angle perceived in these illusions is based on the integration of the local tilts detected by simple cells which is presumed to be a key mechanism utilized by the complex cells to create our final perception of tilt.

## I.  Introduction

We explain the tilt effect in a family of Café Wall illusion with modified mortar lines called dashed Café Wall illusion [1]. As opposed to the original Café Wall pattern with the continuous, uniform grey mortar lines that separate shifted rows of black and white (B/W) tiles [2], in the modified versions there are small dashed lines instead. The dashes are superimposed on tiles with two shades of grey (G), resulting in inducing tilt/bow effects [3]. Based on the relative luminance of the dashes with respect to the tiles, this family is categorized into two major classes of 'contrast' or 'similar' polarity types. Two samples of this family are shown in Fig. 1 which are cropped samples selected from Todorovic's dashed versions of Café Wall [3], to make them symmetrical for our investigations. The left pattern in Fig. 1 is called the Contrast Polarity (CP) version based on the contrast polarity of the dashes with the tiles, and the right pattern shows a 'Similar Polarity' (SP) type. The arrangement of tiles in these two stimuli was also modified compared to the original Café Wall pattern, and due to this, the inducing tilt effects appear as concave/convex bows not divergent/convergent tilts. Although the dashes are arranged horizontally, we encounter reversed tilt orientations (concave vs convex bows) when comparing any two corresponding intersections of rows of tiles in these stimuli. The tilt effect is more dominant in the CP version compared to the SP and we are going to investigate this difference here.

One problem with current vision systems, and in particular those trained by Deep Neural Nets, is that they do not find natural features.  This not only means they are not comprehensible to humans, but that they tend not to translate well to different situations or environments, and furthermore, they are easily spoofed by an adversarial algorithm that makes enough small textural changes to pass threshold on the artificial features so that a computer misinterprets what a human sees as indistinguishable from the original [4]. This work emphasis on processing the way humans do and seeing the features people see, as a bias for computer vision/learning systems.

In Section II we explain the details of a non-classical receptive field (nCRF) model proposed by Todorovic [3] and a list of other models for these types of illusions. We then move to the Vis-CRF[1] [5-9] model used in our study (Section III), before comparing the results of our model with Todorovic's neural image as his model output in Section IV.

## II.  Explanatory models

Todorovic [3] proposed a computational model for explaining the illusory tilts in these so-called "orientation illusions". He noted that the direction of the illusory offset is reversed and appears to be less salient in the SP version. To explain this, he used a non-classical receptive field (nCRF) model based on two odd symmetric receptive fields simulated by differences of offset from elongated Gaussians, tuned to the horizontal orientation. The complementary sensitivity polarities for the receptive fields that show the reaction of the model for two dashed Café Wall illusions (SP and CP versions) were modeled in his work. Todorovic's model outputs are presented with color-coded dots illustrating the activation of units with the corresponding receptive field polarities. The activity levels are shown using dot sizes [3; Figs 3-5] (Please refer to the original work for further details).

---

[1] The Classical Receptive Field (CRF) model for simulating simple cells in our Vision (Vis)

In his work, the output is a neural image that refers to the "alternating elongated streaks of activity of clusters of neurons of the same type" (reaction patterns) [3]. Todorovic made an "aggregate neural image" by summing over a sample of all orientations (sum of all neural images with receptive field's orientations tuned from -90° to +85° with incremental steps of 5°) and showed that the oblique streaks are prominently presented in the aggregate neural image in the CP version as shown in Fig. 4 of his work but not in the SP type [3]. We will examine and generalize the specific images and compare the output of our Vis-CRF model with Todorovic's results in the experimental result section (IV).

There are other explanations for the occurance of illusions similar to the dashed Café Wall illusion such as 'Contrast polarity' and 'Boundary completion' [10-12] of the interacting/elementary units in the stimuli. The majority of these techniques are mid to high-level models for addressing the misperception of tilt orientation. The details of these models are out of the scope of this work since the focus of research here is on receptive field (RF) modeling of the tilt effect. Our model is a simple demonstration for the neural mechanism behind these types of higher-level explanations.

We make explicit the hypothesis that the emergence of local tilt cues in these patterns and Tile illusions originates in the retinal/cortical simple cells (group activations of neighboring simple cells), however, this does not preclude the involvement of more complex cells with the orientation selectivity properties at later stages.

## III. METHOD

The features of Vis-CRF model [5-9] used in our studies (DoG-based) have the characteristics of mammalian early visual processing. In the numerous physiological studies, we find a diverse range of the receptive field types and sizes inside the retina, resulting in a multiscale encoding of the visual scene [13-15]. It has been shown by neuroscientists that the activations of these types of simple cells can be modeled based on Differences and/or Laplacian of Gaussians (DoG, LoG). For modeling the receptive fields of Retinal Ganglion Cells (RGCs), DoG filtering [16, 17] is a good approximation of LoG, if the scale ratio of the surround to the center is close to 1.6 [18, 19] (1.6 ≈ φ, the ubiquitous *Golden Ratio*).

The Difference of Gaussian (DoG) filter used in our Vis-CRF model is given in (1), in which $x$ and $y$ indicating the distances from the origin in the horizontal and vertical axes respectively, where $\sigma_c$ is the standard deviation/scale of the center Gaussian; and $\sigma_s$ is associated with the surround Gaussian. By applying the DoG filters at multiple scales to an image, an intermediate representation for the image is generated called an edge map. Therefore, to model the RGC responses, the DoG transformation creates an edge map representation at multiple scales referred to as EMap-DoG for a given pattern. We further note that the DoG filters are normalized and concentric in Vis-CRF model.

$$DoG_{\sigma_c, \sigma_s}(x,y) = 1/2\pi(\sigma_c^2)\exp[-(x^2+y^2)/(2\sigma_c^2)] - \quad (1)$$
$$1/2\pi(\sigma_s^2)\exp[-(x^2+y^2)/(2\sigma_s^2)]$$

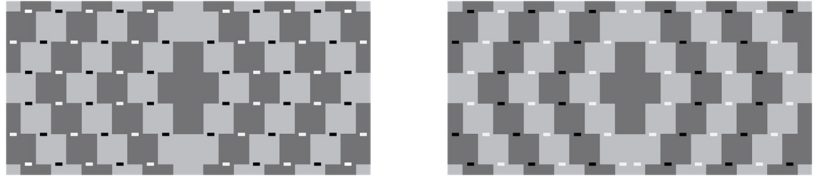

Fig. 1. Two cropped samples from Todorovic's dashed-versions of Café Wall illusion [3]. Left: Contrast Polarity (CP) version, since the overlayed dashes have contrast polarity with the tiles they are positioned on. Right: Similar Polarity (SP) type in which white dashes are positioned on light-grey tiles and black dashes on dark-grey tiles having the same polarity.

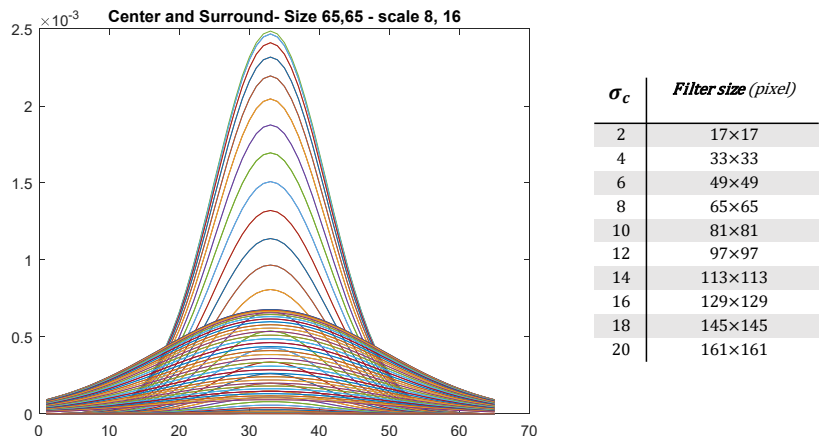

Fig. 2. (Left) 2D surface view of a sample center and surround Gaussians to be used for generating a DoG filter (at $\sigma_c = 8$ and $\sigma_s = 16$). (Right) The relationship between the scale of the center Gaussian ($\sigma_c$) and the *Window ratio* (*h*) in the DoG-based model (Vis-CRF). The range of DoG scales is defined based on the stimulus features.

$$s = \sigma_{surround} / \sigma_{center} = \sigma_s / \sigma_c \qquad (2)$$

The scale ratio of the surround to the center in the DoG filter is determined by a parameter referred to as *Surround ratio* ($s$) in our model given in (2). We typically use $s = 2.0$ or $1.6$ in the model because the ratio of 1:1.6 – 2.0 (indicating the size of the center: surround Gaussians) is a typical range for modeling simple cells. (1:1.6 is given by Marr and Hildreth [18] for modeling retinal GCs in general, and Earle and Maskell [19] used this ratio for DoG modeling specifically to explain the Café Wall illusion). Other ratios have been used by other researchers, for example, the ratio of 1:5.0 in Lulich and Stevens [20]. We reiterate that Vis-CRF model is insensitive to changes in center-surround ratio in the range tested from 1:1.4 to 1:8.0. Our model with its 1.6-2× ratio relates to simple cells (for example midget bipolar cells), whereas the ratio 4-8× relates to complex cells [21 - p. 17090].

In respect to the filter size, the DoG is only applied within a window in which the values of both Gaussians are insignificant outside the window. We defined a parameter called *Window ratio* ($h$) to control window size in the model as given in (3):

$$Window\ size = h \times \sigma_c + 1 \qquad (3)$$

Parameter $h$ determines the outline of the DoG filter by showing how much of each Gaussian (center and surround) is included inside the filter. In particular, $h = 2$ corresponds to DoG filters containing 68% center, 31% surround; $h = 4$ corresponds to filters with 95% center, 68% surround; and $h = 8$ corresponds to the standard p<0.05 significance for the surround Gaussian that is 99.94% center, 95% surround. $h = 8$ is used in the experimental runs.

Fig. 2 (Left) illustrates 2D graphs of separate center and surround Gaussians with their difference giving the DoG filter. The connection of Window ratio ($h$) to the shape of the DoG filter ($\sigma_c$) has been tabulated in the right of the figure. In Vis-CRF model, we examined varying scale ratios and window size/ratio of Gaussian filters. This approach is innovative as it addresses the shortcoming of other studies that fail to emphasize the importance of the choice of these parameters, especially the second one ($h$), contributing to the performance of their respective filter/models. The range of the DoG scales is defined in such a way as to capture both high-frequency details of edges and textures as well as the low-frequency profile conveyed by brightness/color from the objects within the scene.

## IV. Experimental Results

The edge map in the model is similar to Marr's low-level representation [18] of visual input and differs from the typical approaches of the DoG or LoG based pyramids such as SIFT or SURF [22-24]. To thoroughly investigate the emergence of tilt cues we need to analyze the edge map at multiple scales which is a layered filtered response of the visual content. However, the DoG/LoG pyramid models attempt to compute a weighted sum of all the filtered outputs (from different scales) to present a multiscale representation as an output estimation for the later processing stages done by more complex cortical cells.

Therefore, in our DoG implementations, we tune the sigmoid to match the resolution of the image features. The size and resolution of the image features and the model predictions across different sizes and resolutions have been reflected in our previous papers for a variety of Tile illusions [6-8, 25]. The stimuli used here have spatial dimensions of 1345×630px for the CP version and 1354×630px for the SP stimulus (Fig. 1). The tiles are approximately 112×112px and the width of dashed lines is around 10px. Based on the feature sizes in the stimuli, the range of DoG scales are chosen empirically between $\sigma_c = 2$ to $\sigma_c = 20$ with incremental steps of 2 to extract essential visual information out of these patterns.

Fig. 3 shows the DoG edge maps at ten scales for these two stimuli. The edge maps are presented in 'black and white' and using the 'jetwhite' false-color map [26] with color bars next to them. For our explanations, we focus on the black and white edge maps while the false-color images provide further detail that will allow the reader to develop additional insight into what is happening here. Based on the outputs, we do not see much difference in the edge maps of these stimuli at scale 2 ($\sigma_c = 2$). However, when the scale increases, we can see some substantial changes in the groupings of tiles with the dashed mortar lines that start to emerge around medium scales ($\sigma_c = 8, 10$). From the finest scale to scale 10, the groupings of the same colored tiles in both stimuli follow nearly the same pattern although some minor differences can be found on the appearance of the dashed lines in the edge maps and also in the interior of the tiles.

In the CP stimulus from scales 12 and 14 upwards, the directions of groupings of identically colored tiles in the edge maps are changed that is incompatible with the previous groupings of tiles at finer scales. This does not occur for the SP stimulus where the grouping stays the same across multiple scales of the edge maps. In the CP version, the tilt cues for bowing start to appear at medium scales ($\sigma_c = 12, 14$) of the edge maps, revealing the underlying neural activations associated with our final tilt perception in these stimuli. The tilt cues that emerge in the edge maps across these scales are consistent with our perceptual grouping of tiles in these variations and this seems to be the preliminary basis of the way we perceive these illusory tilts. What we have found from the broad range of investigations on the edge maps for Tile illusions, in general, is that if the edge map reveals two incompatible groupings of pattern elements across different scales (tiles in these stimuli), then we encounter tilt or distortion illusion similar to what we can see in Fig. 3 for the CP version.

In previous work [6-8], it was noted that for detecting near-horizontal tilted line segments in the Café Wall illusion known as the twisted cord elements [19, 27, 28] in the edge maps, the DoG scale should be close to the mortar size (here the dashed widths) and this is the same for these stimuli. We also saw that "persistence of mortar cues" across multiple scales plays a major role in determining how strongly we perceive the induced tilt effect in Café Wall illusions [7]. This can be seen clearly here for the CP version with more persistent local tilt cues across multiple scales of the edge maps till scale 18, but a few scales lower for the SP type (at scale 12 or 14). This is consistent with our perceptual tilt effects in these stimuli with a stronger tilt effect in the CP type.

Often these patterns/stimuli are made up by hands that can introduce subtle errors with a possible enhancement to the illusion, whereas programmable solutions that can be produced by software can be used instead to validate the image. Rather than the two samples uses here as cropped stimuli from Todorovic's article [3] for the sake of a clear comparison between these two methods, the rest of the stimuli investigated by the model are all computationally generated images in MATLAB that let us adjust the characteristics of image feature such as contrast, scales, phase shift and many more. This also eliminates any additional tilt cues/ subtle errors in the generated patterns and guides us through the investigation of the main factors contributing to the tilt effects in geometrical distortion stimuli (Tile illusions).

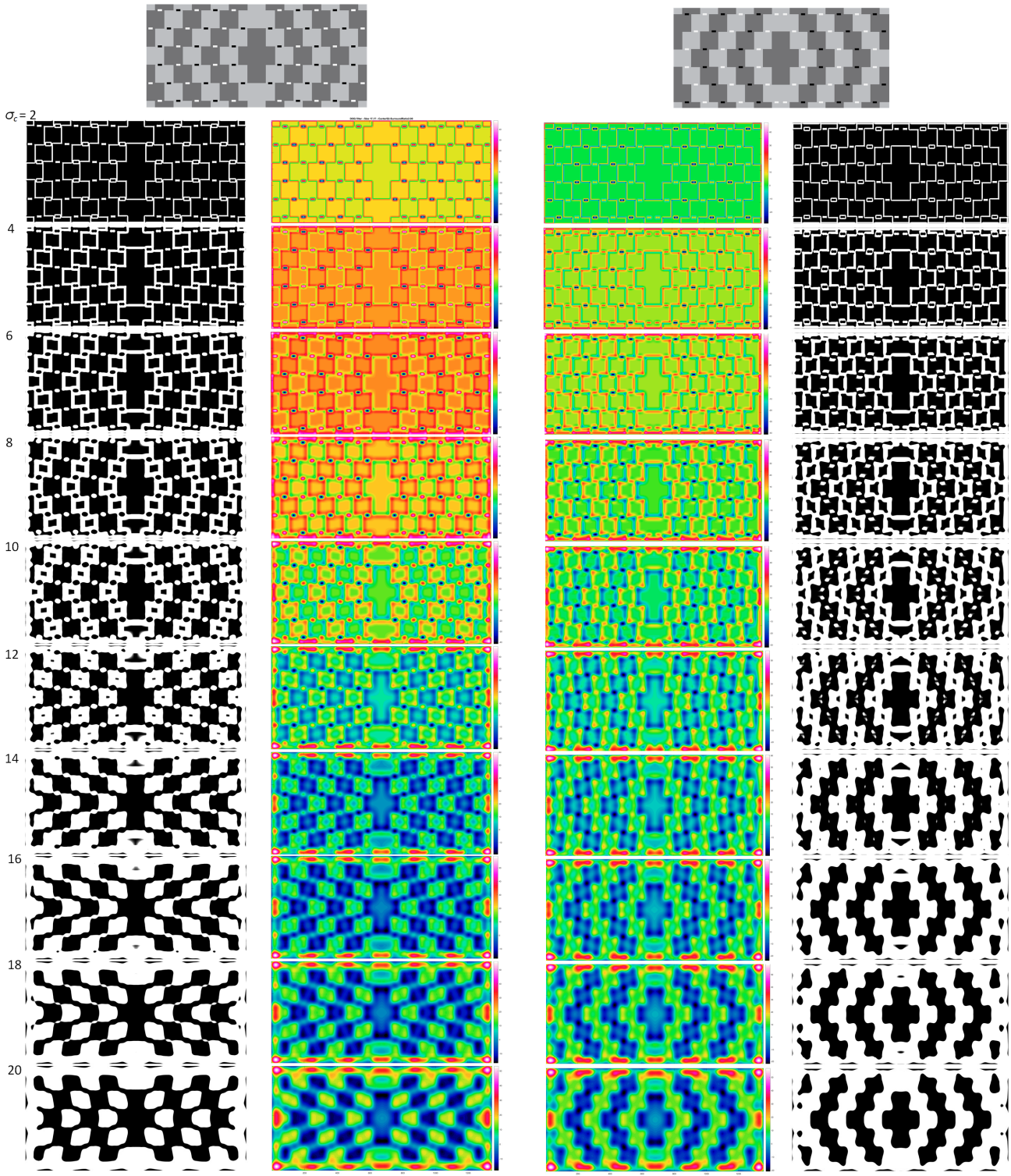

Fig. 3. The DoG edge maps at 10 different scales ($\sigma_c$ = 2 to 20 with incremental steps of 2) for the CP (on the left) and SP (on the right) samples of dashed Café Wall illusion, shown in both black and white as well as in jetwhite color map [26].

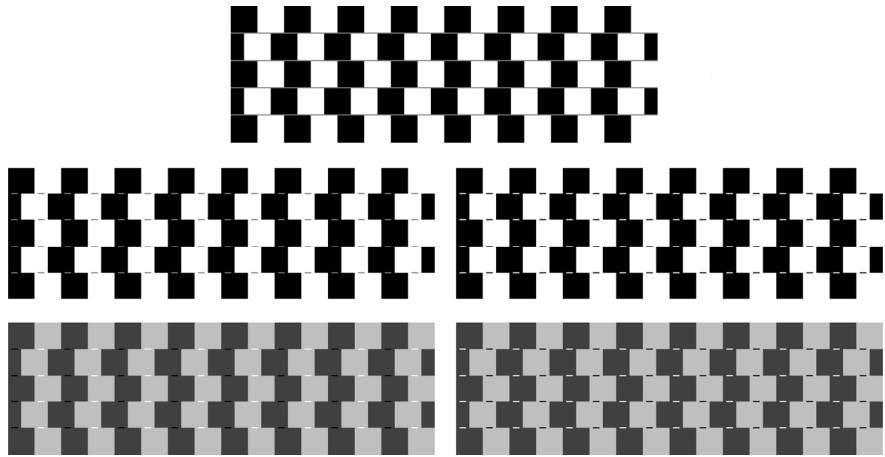

Fig. 4. Café Wall variations generated by Matlab with a tile size of 200×200px and mortar width of 8px (same widths for dashed lines). Top: Original Café Wall pattern. Middle: Simple Dashed version of the original Café Wall pattern (B/W tiles with grey dashes) on the left and the Contrast Polarity (CP) version for B/W tiles on the right. Bottom: Similar polarity (SP) type on the left and Contrast Polarity (CP) version on the right for grey (G) tiles with two shades of grey.

Fig. 4 (Top) shows the original Café Wall pattern [2] designed with a spatial resolution of 200×200px tiles and 8px mortar. In the second row, a simple dashed version with dashes of an intermediate grey luminance between the B/W tiles is shown on the left. Then a dashed version of type CP with B/W tiles is shown in the middle-right. At the bottom of the figure, SP and CP versions are shown for grey (G) tiles, with similar characteristics to the Todorovic's versions. We perceive stronger tilt effects in the CP versions, presented on the right side of the figure (for the B/W as well as the G tiles).

Two more samples of the dashed Café Wall illusions with similar inducing bow effects to the Todorovic's variations are shown in Fig. 5. Both of these patterns are symmetrical (both horizontally and vertically) and generated by concatenating a Café Wall pattern with its vertical mirrored image and then with the horizontally mirrored image from the previous stage. In the CP version, we see two inducing bows towards the center in the middle and then an opposite direction of the bow going outwards to the following intersections of rows of tiles (the bows are interchangeable in each intersection of rows from concave to convex and vice versa). For the SP version (Fig. 5- Right), we see opposite outgoing bows from the center in the middle of the pattern and similar to the CP version, interchanging the bow's direction in the following intersections going away from the center. In the SP version, the inducing tilt/bow is not as dominant as the CP version on the left, but it is still tractable when comparing these two patterns together. The tilt effect becomes more conspicuous when we take our focus away from the center to the periphery of the pattern or simply by shifting our gaze away from their centers.

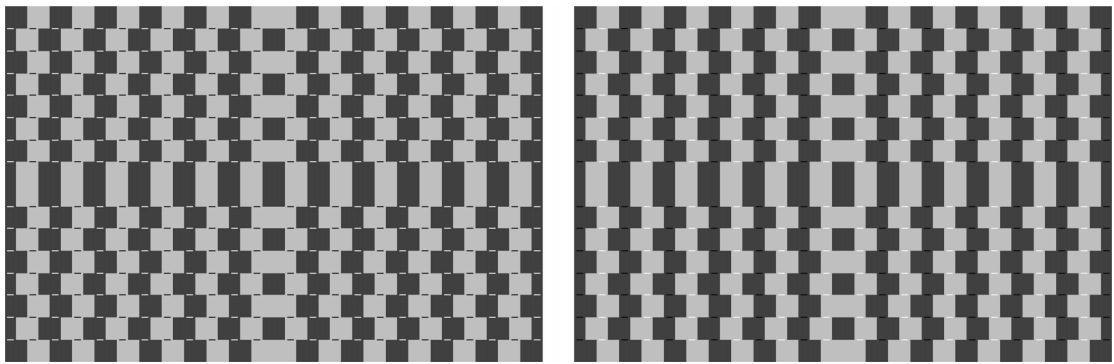

Fig. 5. Two samples of the dashed Café Wall illusion designed by the author which have similar inducing tilt effects to the Todorovic's variations but without any additional tilt cues.

We treat the dashed Café Wall patterns investigated here as an example and this classical Gaussian Receptive Field model (Vis-CRF) can be used as a generalized model for predicting different illusory tilts for other Café Wall patterns with varying parameters. These include patterns with different dash-sizes, dash-contrast, and phase shifts for their positions from the intersections of rows of tiles, even dot-versions of the Café Wall illusion [11, 12] that can be investigated similar to the wide range of Café wall patterns studied in our previous work [7-9, 29].

## V. CONCLUSIONS

We have successfully demonstrated that a simple Classical Receptive Field (CRF) model implemented by isotropic Differences of Gaussian (DoG) filters for simple cells in early stages of vision, with the DoG scales tuned to the object/texture sizes in the pattern, can explain the emergence of tilt and distortion illusions in this family of Café Wall illusion. Similar conclusions have been drawn for other variations of the Café Wall illusions, reflected in our previous work [6-8, 29]. A critical point to note here is that although Todorovoc [3] explained the tilt effect in these stimuli based on a nCRF model with a diverse range of orientations for the elongated Gaussians, the effect of the scale for the filters in his explanations was completely ignored, as he just considered one scale (noted to be the resolution of the filters). The main contribution of this work is to highlight the most critical factors/mechanisms involved in our low-level vision for revealing the illusory tilt cues in these patterns based on the processing priorities from the retina to the cortex. We have shown here that a simplified isotropic DoG filter at multiple scales, modeling lateral inhibition of simple cells, can encode the illusory tilt information of these patterns. It seems that more complex cells will participate in higher-level processing stages for the final tilt perception such as for estimating the angle of tilt in the edge map. We highlighted the fact that it may be too early to reject concentric filters all together in favor of elongated versions, modeling the orientation selectivity properties of more complex cells, for explaining a mid-/high-level visual phenomenon. Todorovic acknowledges that similar results to his model can be achieved "with even symmetric and circular (concentric-antagonistic) receptive fields" [3], and that "the exact orientation tuning of the members of these populations may be of secondary importance" [3] and our simulation results are a proof for this.